\documentclass{iau}
\usepackage{natbib}
\usepackage{graphicx}
\usepackage{aas_macros}

\title[JD 11.~~Collisionless Dynamics and the Cosmic Web] 
{Collisionless Dynamics and the Cosmic Web}

\author[O. Hahn]   
{Oliver Hahn}

\affiliation{
Department of Physics, ETH Zurich, \\CH-8093 Z\"urich,
  Switzerland \\ email: {\tt hahn@phys.ethz.ch} 
}

\pubyear{2014}
\volume{308}  
\pagerange{???--???}
\jname{The Zeldovich Universe}
\editors{R. van de Weygaert, S. Shandarin, E. Saar \& J. Einasto, eds.}
\begin{document}

\maketitle

\begin{abstract}
I review the nature of three-dimensional collapse in the Zeldovich approximation, how it relates to the underlying nature of the three-dimensional Lagrangian manifold and naturally gives rise to a hierarchical structure formation scenario that progresses through collapse from voids to pancakes, filaments and then halos. I then discuss how variations of the Zeldovich approximation (based on the gravitational or the velocity potential) have been used to define classifications of the cosmic large-scale structure into dynamically distinct parts. Finally, I turn to recent efforts to devise new approaches relying on tessellations of the Lagrangian manifold to follow the fine-grained dynamics of the dark matter fluid into the highly non-linear regime and both extract the maximum amount of information from existing simulations as well as devise new simulation techniques for cold collisionless dynamics.
\end{abstract}

\firstsection 
\section{Introduction}
\begin{figure}[t]
\begin{center}
 \includegraphics[width=3in]{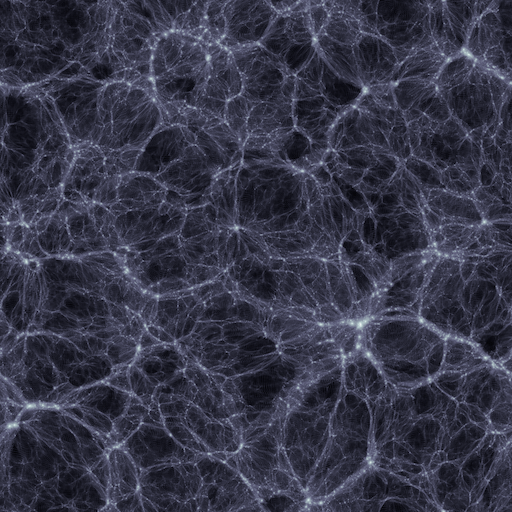} 
 \caption{The cosmic web as formed in an N-body simulation of the $\Lambda CDM$ cosmology at $z=0$. The width of the image is $250\,h^{-1}{\rm Mpc}$, shown is the logarithmic overdensity $\log(1+\delta)$. Zel'dovich's first order perturbation theory predicts the formation of pancakes, filaments, halos and voids in the cosmic matter distribution. Since in CDM perturbation exist down to very small scales, filaments are of course made up of haloes, and pancakes of filaments.}
   \label{fig1}
\end{center}
\end{figure}

\noindent Zeldovich's legendary formula \citep{Zeldovich1970} describes cosmological structure formation and in particular the emergence of singularities under gravitational collapse that emerge from critical curves in the velocity perturbation field. It arises rather simply from the Lagrangian motion of pressure-free fluid parcels under self gravity. The relation between Lagrangian and Eulerian space can at all times be described by the position $\mathbf{x}$ and velocity $\mathbf{v}$ of a fluid parcel labeled by its three-dimensional Lagrangian coordinate $\mathbf{q}$ as formally given by
\begin{equation}
\mathbf{x}_\mathbf{q}(t) = \mathbf{q} + \mathbf{L}(\mathbf{q},\,t),\quad\textrm{and}\quad\mathbf{v}_\mathbf{q}(t) = \dot{\mathbf{x}}_\mathbf{q} = \dot{\mathbf{L}}(\mathbf{q},\,t).
\end{equation}
The displacement field $\mathbf{L}$ at linear perturbative order is simply proportional to the growing mode of the initial velocity perturbation field. It evolves with the growth factor of linear density perturbations
\begin{equation}
\mathbf{L}(\mathbf{q},\,t)\propto D_+ \mathbf{v}_\mathbf{q}(0),\quad\textrm{and}\quad\dot{\mathbf{L}}(\mathbf{q},\,t)\propto \dot{D}_+ \mathbf{v}_\mathbf{q}(0),
\label{eq:zeldovich_approx}
\end{equation}
which {\em is} the Zeldovich approximation, implying that the fluid elements move on straight lines as determined by their initial velocity vectors. The density then becomes obviously (simply through the Jacobian of the transformation from $\mathbf{q}$ to $\mathbf{x}$)
\begin{equation}
1+\delta = \prod_{i=0}^{3} \left(1+{\rm D}_+ \lambda_i\right)^{-1} , \quad\textrm{where}\quad \lambda_i = {\rm eig}\left\{ T_{ij}\equiv\partial_{ij} \phi\right\},
\label{eq:density}
\end{equation}
and $\phi$ is the velocity potential (i.e. $\mathbf{v_q}(t)\propto\dot{D}_+ \boldsymbol{\nabla}\phi$), which is, at first order, proportional to the gravitational potential. Since $T_{ij}$ is a symmetric and real tensor, its eigenvalues are real numbers and thus can be arranged $\lambda_1\le\lambda_2\le\lambda_3$. This however implies that $\delta$ can undergo a maximum of three singularities over time, depending on the number of negative eigenvalues $\lambda_i$. If $\lambda_1 < 0$, then at some finite time $t_1$, when $D_+(t_1)\lambda_1 = -1$, the fluid element will undergo a singularity along the eigenvector associated with $\lambda_1$, followed by the other axes assuming they correspond to negative eigenvalues. Thus, the first objects that form are two-dimensional structures (the infamous Zeldovich pancakes). Then the collapse along a second axis happens and a one-dimensional structure emerges (a filament), before also the third axis collapses and finally a (roughly) spherical structure forms. Of course this is a somewhat simplistic picture and in general the singularities/catastrophes that can arise have a multitude of structures (see \citealt{Hidding2014} for an extensive discussion of catastrophes in the Zel'dovich approximation). Since density perturbations exist on various scales, this pancake collapse scenario occurs simultaneously on different scales leading to the multi-scale nature of the cosmic web. On the smallest scales sit the halos, embedded in larger perturbations for which the final direction has not collapsed yet, corresponding to filaments. In turn, the filaments are embedded in even larger scale perturbations for which two axes have not collapsed yet, corresponding to the pancakes. Naturally, if any eigenvalue is smaller than zero, the corresponding axis will never collapse in this approximation. In Figure~\ref{fig1}, we show the cosmic web as it emerges in a density map of an $N$-body simulation of a $\Lambda$CDM cosmology. 

As is well known, the approximation of eq.~(\ref{eq:zeldovich_approx}) breaks down in two cases: (1) after shell-crossing in one-dimension, when the gravitational force acting on a fluid parcel reverses its direction, and (2) in multiple dimensions if higher derivatives of the gravitational perturbation field (e.g. $\partial_{ijk}\phi$) become non-negligible compared to the tidal (i.e. $\partial_{ij}\phi$) term \citep[e.g.][]{Crocce2006}. We will give point (1) more attention next.

\begin{figure}
\begin{center}
 \includegraphics[width=2.7in]{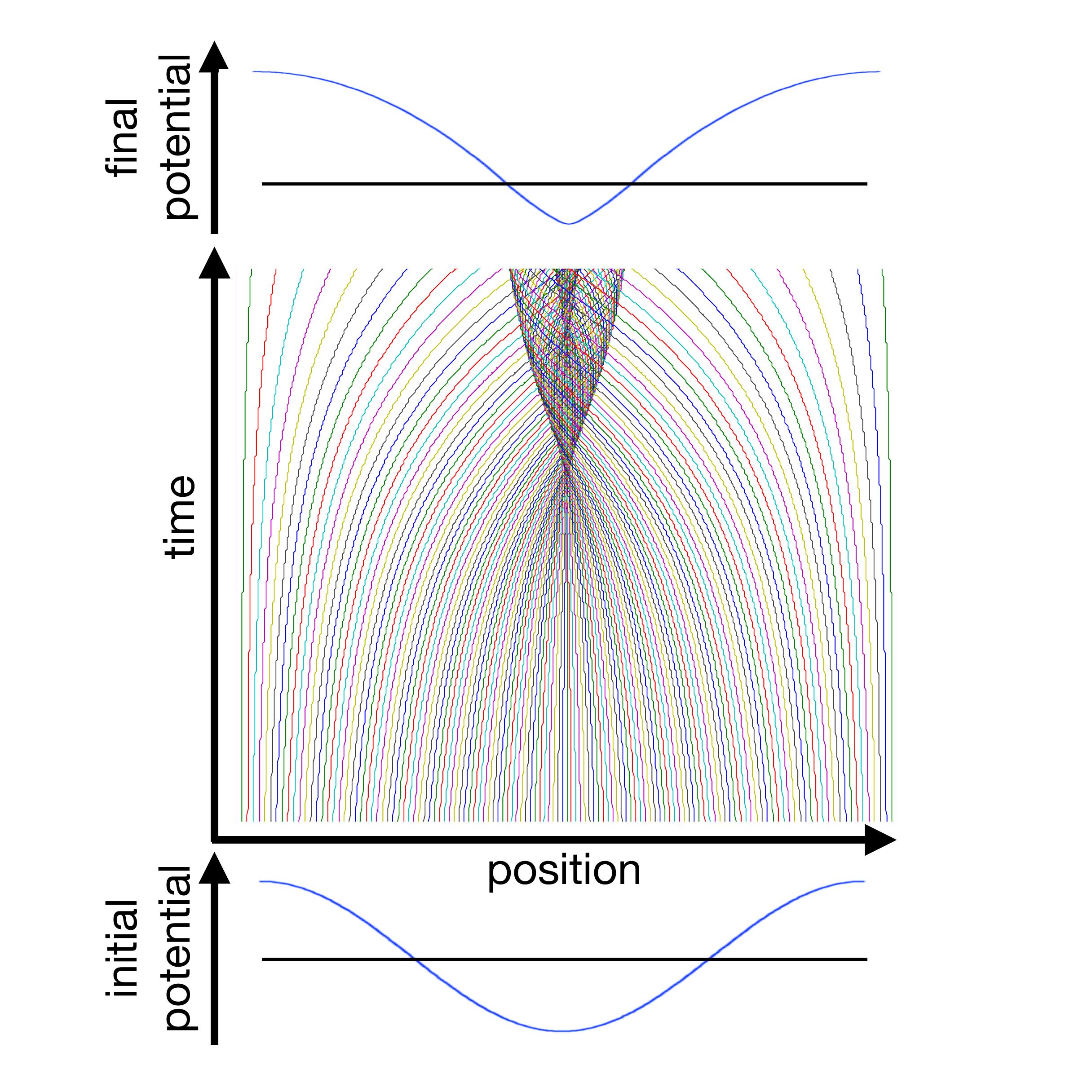} 
 \includegraphics[width=2.3in]{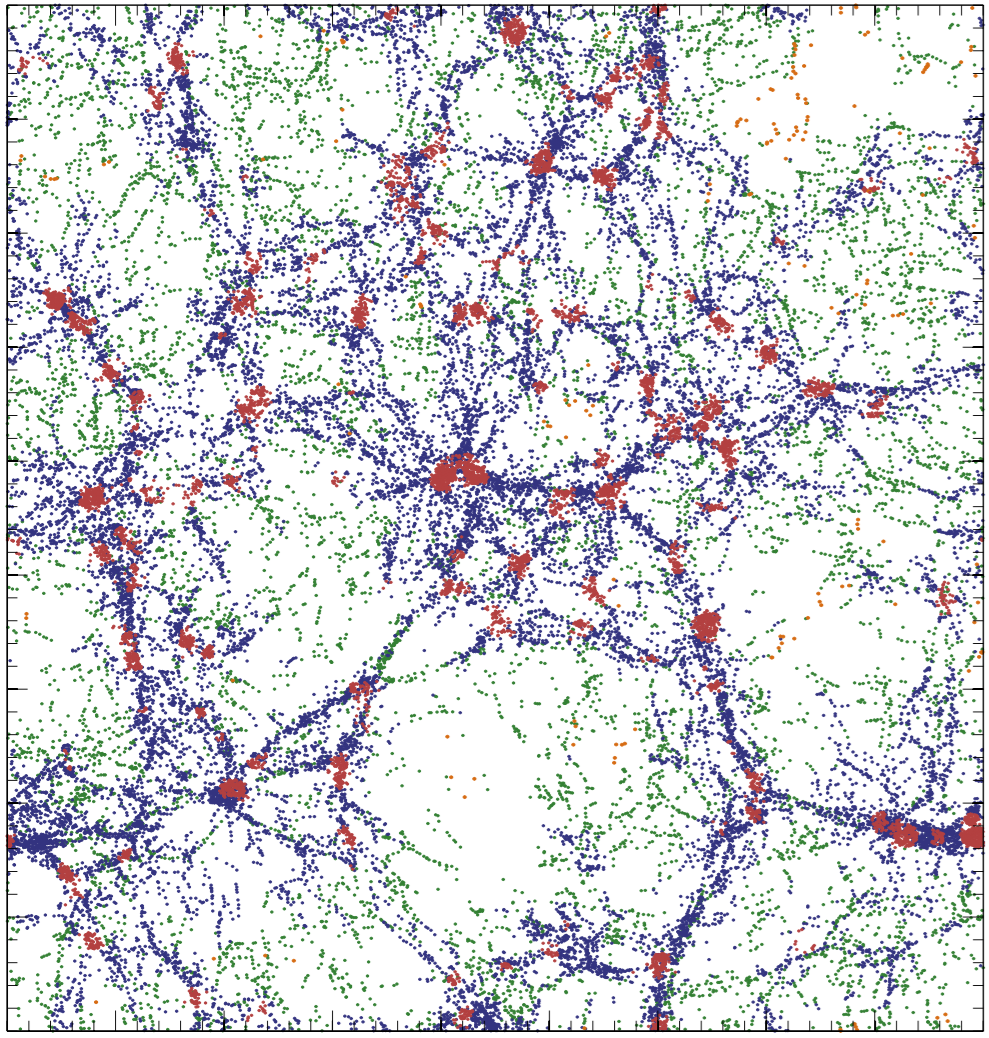} 
 \caption{{\bf Left:} Collapse of a plane wave under self gravity. Shell-crossing leads to a gravitationally bound structure forming in the centre. Regions of tidal expansion in the final potential correspond to regions outside the bound structure and regions of tidal compression to the inside. {\bf Right:} Haloes from an $N$-body  $\Lambda$CDM simulation classified by the signature of the tidal tensor eigenvalues smoothed on $2\,h^{-1}{\rm Mpc}$ splitting them into clusters (red: -,-,-), filaments (blue: +,-,-), walls (green: +,+,-) and voids (yellow: +,+,+) according to the method of \cite{Hahn2007a}. The image is $180\,h^{-1}{\rm Mpc}$ wide. }
   \label{fig2}
\end{center}
\end{figure}

\section{The Zeldovich approximation, non-linear structure formation and identification of dynamically distinct structures of the cosmic web}
\noindent In Figure~\ref{fig2}, we show the collapse of a collisionless self-gravitating plane wave, i.e. the collapse of a sinusoidal perturbation $\mathbf{L}(\mathbf{q},0) \propto  \sin(\mathbf{k}\cdot \mathbf{q})$, or equivalently $\phi \propto -\cos(\mathbf{k}\cdot\mathbf{q})$, where $\mathbf{k}$ is the wave vector of the perturbation. We plot the fully non-linear solution in a slightly different way than is done usually. The bottom panel shows the initial gravitational potential, the middle panel shows the trajectories of particles over time (a given curve corresponds to a fixed $\mathbf{q}$), and the top panel shows the gravitational potential at the final time. Time is plotted logarithmically in the middle panel to highlight better the dynamics around the time of first shell-crossing. Several important observations can be made from this simple diagram, which is perfectly reproduced by the Zeldovich approximation for all points outside of the outer caustics of the multi-stream region. In the fully non-linear setting, repeated shell-crossing happens in the center due to self-gravity. We observe that (1) the initial tidal field is given by $\lambda=\partial_{ij}\phi \propto \cos(\mathbf{k}\cdot\mathbf{q})$ implying that trajectories in the central region are convergent ($\lambda<0$), and trajectories outside are divergent ($\lambda>0$). Of course, the outer trajectories still collapse onto the central structure, but the Eulerian volume which had $\lambda>0$ initially remains divergent and becomes underdense. (2) the final potential has not changed qualitatively from the initial. Obviously, it still has a convergent ($\lambda<0$) central part and divergent ($\lambda>0$) outer parts (in both cases separated by the horizontal black lines which are drawn to intersect at the inflection points of $\phi$). Furthermore, the convergent region agrees well with the shell-crossed region. This latter observation provides a strong motivation to use the final tidal field to classify the cosmic web into dynamically distinct regions as suggested by \cite{Hahn2007a}.

Naturally, in CDM, almost all matter is in dark matter halos so that a simple classification by the final potential would identify halos as dips in the potential rather than describe the large-scale structure. For this reason, \cite{Hahn2007a} smoothed the potential on about twice the virial scale of $M_\ast$ halos of $\sim2\,h^{-1}{\rm Mpc}$ at $z=0$ and on correspondingly smaller scales at high redshift \citep{Hahn2007b}. The resulting smoothed gravitational potential $\psi$ can then be classified into distinct regions according to the signature of the eigenvalues of $\partial_{ij}\psi$, just as in eq.~(\ref{eq:density}). Regions of three-dimensional compression $(-,-,-)$ correspond then to tidal compression on the smoothed scale and thus can be identified with clusters (or nodes of the web), and corresondingly $(+,-,-)$ with filaments, $(+,+,-)$ with walls and $(+,+,+)$ with voids. The performance of such a classification can be seen in the right panel of Figure~\ref{fig2}. Very clearly, the method is able to identify the nodes of the cosmic web and the larges filaments very well and assigns the finer filaments to walls. However, One can immediately see from this classification that the volume assigned to voids is very small, which subsequently has led \cite{Forero2009} to argue that the signature should not be evaluated w.r.t. to zero as this describes only the asymptotic behavior, i.e. whether $1+D_+\lambda$ vanishes within a finite time in eq.~(\ref{eq:density}). They have thus introduced an additional (free) parameter $\lambda_{th}$ with respect to which the eigenvalues are compared, i.e. the signature of $(\lambda_i+\lambda_{th})$ is determined instead. Another (arguable) shortcoming is the fixed smoothing scale, which e.g. identifies all nodes of the cosmic only on one scale. The combination of multi-scale filtering with an evaluation of the tidal eigenvalues has been discussed by \cite{Cautun2013}. 

During the linear stages of structure formation -- as expressed by the Zeldovich approximation -- the velocity field and the gradient of the gravitational potential are proportional to each other (cf. eq. \ref{eq:zeldovich_approx}). This implies that at early times a classification by the eigenvalues of the tidal tensor ${\rm eig}\, \partial_{ij}\psi$ is equivalent to a classification by the velocity deformation tensor ${\rm eig}\, \partial_i v_j$. At late times, this equality is however broken by non-linear terms so that \cite{Hoffman2012} have proposed to use the latter when classifying structures of the cosmic web. In particular, while primordial vorticity is usually neglected since it is a decaying perturbation, at late time vorticity arises in the non-linear cosmic velocity field (see the discussion in the next section). In addition to these approaches inspired by the evolution of large-scale structure in the Zeldovich approximation, also density fields and logarithmic density fields have been used \citep[most notably][]{Aragon2007,Aragon2010,Sousbie2011,Cautun2013}. With such a multitude of classifiers available, the question what needs to be classified arises, since all definitions are to a large degree arbitrary. What all methods have achieved is to provide a clear prediction for the alignment of halo spins and shapes with the cosmic web \citep[e.g.][among others]{Aragon2007a,Hahn2007b,Codis2012}. It still remains to be seen whether there are other observable properties of galaxies that depend on the cosmic web environment rather than just the halo mass and the central vs. satellite distinction.

\section{The Zeldovich approximation, the Lagrangian manifold and Lagrangian tesselations}
\noindent Another interesting property of the mapping between Lagrangian and Eulerian space is that it has a manifold structure and directly describes the distribution function of a perfectly cold fluid \citep{Arnold1982,Shandarin1989}. This means that the mapping between Lagrangian space and Eulerian phase space varies smoothly between points close in Lagrangian space, i.e. the mapping is differentiable,
\begin{equation}
\mathbf{q}\mapsto(\mathbf{x_q}(t),\mathbf{v_q}(t)),\quad\Rightarrow\quad T_\mathbf{q}=\left(\boldsymbol{\nabla}_\mathbf{q}\mathbf{x},\boldsymbol{\nabla}_\mathbf{q}\mathbf{v}\right),
\end{equation}
where the latter defines the three-dimensional space tangent to the cold distribution function in six-dimensional phase space. If the cold distribution function is initially sufficiently smooth, it will remain so by virtue of the collisionless Vlasov-Poisson system of equations. The density of dark matter in the vicinity of $\mathbf{q}$ is given by the determinant of the spatial part, i.e. 
\begin{equation}
\rho = m_p \left| \det \frac{\partial x_i}{\partial q_j} \right|^{-1},\label{eq:density2}
\end{equation}
where $m_p$ is the particle mass. Notably, as we had already seen in eq.~(\ref{eq:density}), the density becomes singular in caustics, where this determinant vanishes, i.e. where the distribution function has a tangent perpendicular to configuration space. The evolution of the tangent space can be described in terms of a geodesic deviation equation around point $\mathbf{q}$ \citep{Vogelsberger2008} and allows one to study in simulations e.g. how often a particle goes through a caustic \citep{White2009}. 

\begin{figure}[t]
\begin{center}
 \includegraphics[width=\textwidth]{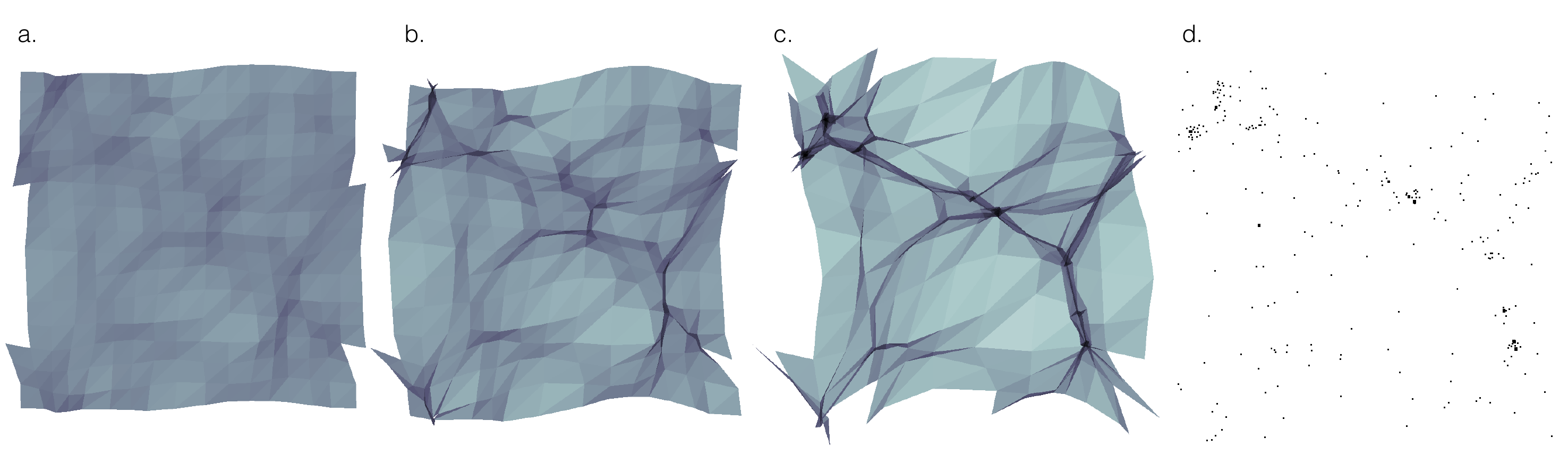} 
 \caption{Foldings of the triangulated phase space sheet in 2+2 dimensional phase space over time. Panels (a) to (c) show the time evolution of a two-dimensional N-body simulation of only $16^2$ particles where the density field is given by triangles connecting particles neighbouring in Lagrangian space. The resulting density estimate is defined everywhere in space. Panel (d) shows the positions of the N-body particles at the same time as panel (c).}
   \label{fig3}
\end{center}
\end{figure}

More recently however, \cite{Abel2012} and \cite{Shandarin2012} noted that the tangent space can be easily approximated in simulations by combining information from neighboring points in Lagrangian space. This can be achieved by determining the neighbors of point $\mathbf{q}$ by a Delaunay triangulation of the regular lattice that the $N$-body particles form in Lagrangian space. Thus, the discrete points $\mathbf{q}_i$ (i.e. the N-body particles) can be associated with the vertices of a tetrahedral mesh that covers all of Lagrangian space, and thus also all of Eulerian configuration space. Each tetrahedron uniquely defines an approximation to $T_\mathbf{q}$, i.e. one can calculate $\partial x_i/\partial q_j$ and $\partial v_i/\partial q_j$ uniquely for each tetrahedron. This in turn enables one to calculate the density of a tetrahedron according to eq.~(\ref{eq:density2}). This is the single-stream density. Since after shell crossing several tetrahedra may overlap a given point $\mathbf{x}$ in configuration space, the densities of all tetrahedra overlapping that point need to be added up to obtain the configuration space density. We illustrate the procedure in two dimensions in Figure~\ref{fig3}. The left-most panel shows the initial triangular mesh. Before shell-crossing occurs, the triangles do not overlap. Density perturbations come from particles that are displaced between Lagrangian and Eulerian space leading to density perturbations between the triangles consistent with Zeldovich's formula. Over time, the perturbations grow, shell-crossing occurs and several triangles can overlap the same point in configuration space. Clearly the filamentary cosmic web emerges already from this simple demonstration. For comparison, the right panel shows the locations of the N-body particles where the filamentary structure is by far not as clearly visible as in the triangulated density estimate that tracks explicitly the anisotropic local deformation tensors.

The approximation of the phase space sheet by finite volume tetrahedral elements provides access to a wide range of properties of dark matter flows that could not be measured before. Most notably, the tetrahedral elements allow to measure  density fields with significantly reduced noise without the need for any smoothing. This allows for rather breath-taking possibilities for visualization \citep{Abel2012,Kaehler2012} or gravitational lensing simulations with suppressed noise \citep{Angulo2014} and allows to,  e.g., explicitly count the number of streams overlaying points in configuration space \citep{Shandarin2012,Abel2012}. They also provide a new way to determine whether structures are collapsed (and thus folded) along three, two, one or no directions which allows one to more directly identify halos, filaments, walls and voids \citep{Neyrinck2012,Falck2012}.

Furthermore, this Lagrangian tessellation approach provides excellent estimates of cosmic mean velocity fields and their derivatives \citep{Hahn2014}. Since the single stream density is known, a very accurate estimate of the bulk velocity field can be obtained. It is defined as the density weighted mean velocity 
\begin{equation}
\left< \mathbf{v} \right>  = \frac{\sum_{s\in{\rm S}} \mathbf{v}_s(\mathbf{x})\,\rho_s(\mathbf{x}) }{\sum_{s\in{\rm S}} \rho_s(\mathbf{x})},
\end{equation}
where $S$ stands for all streams that contain a point $\mathbf{x}$ in configuration space. This allows one to define the {\em exact} differentials of the collisionless mean multi-stream velocity field as
\begin{equation}
\boldsymbol{\nabla}\cdot \left<\mathbf{v}\right> =
\bigl<
\left(\boldsymbol{\nabla}\log\rho\right) \cdot
\left(\mathbf{v}-\left<\mathbf{v}\right>\right) \bigr> + 
\left<\boldsymbol{\nabla}\cdot\mathbf{v}\right> \label{eq:tet_divergence}
\end{equation} 
and
\begin{equation} 
\boldsymbol{\nabla}\times \left<\mathbf{v}\right>  = 
 \bigl<\left(\boldsymbol{\nabla}\log\rho\right)\times\left(\mathbf{v}-\left<\mathbf{v}\right>\right)\bigr>,\label{eq:tet_curl}
\end{equation} 
which hold if the derivative is not taken across caustics. Most notably, the curl of the mean field vanishes in single stream regions since gravity cannot generate vortical modes, but it can arise in multi-stream regions as a 'collective' phenomenon \citep[see also][]{Pichon1999}. We furthermore see that the velocity divergence also has a 'collective' term, which depends on alignment of local flows with the density gradient, in addition to the single stream divergence. 

\begin{figure}[t]
\begin{center}
 \includegraphics[width=\textwidth]{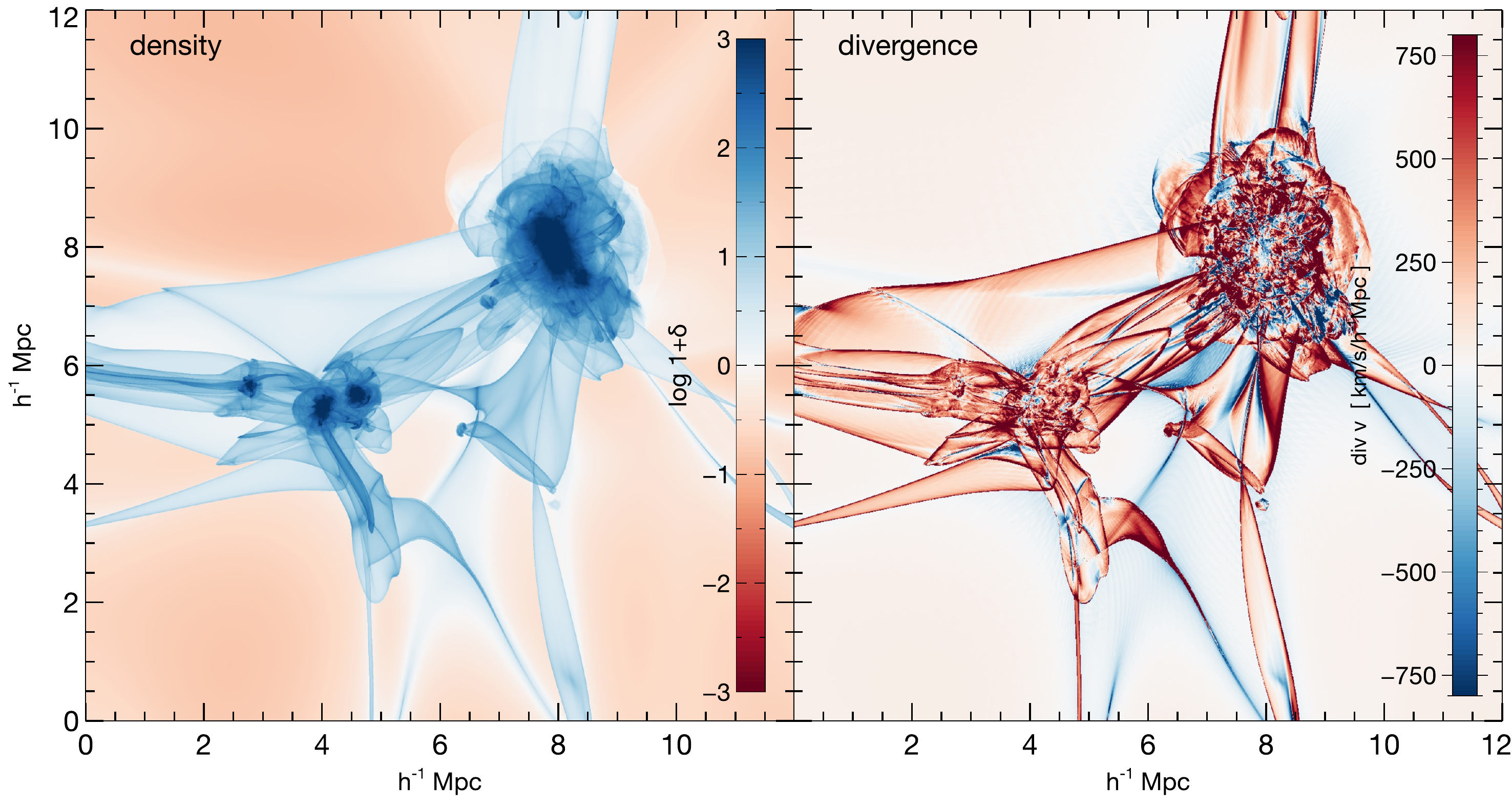} 
 \caption{Slices through the density (left) and bulk velocity divergence field (right panel) from a warm dark matter simulation. Before shell-crossing underdense regions are divergent, and overdense regions are convergent. After shell-crossing, overdense regions are predominantly divergent as well which reflects that the volume of the dark matter sheet is growing over time.}
   \label{fig4}
\end{center}
\end{figure}

In Figure~\ref{fig4}, we show a slice through the density field (left panel) and the velocity divergence field (right panel) calculated from an $N$-body simulation of halo in a warm dark matter cosmology. We see that underdense regions have positive velocity divergence before shell-crossing, while overdense regions have negative divergence. After shell-crossing, overdense regions also show a positive divergence. The correlation before shell crossing follows directly from a series expansion of eq.~(\ref{eq:density}) which yields $1+\delta\simeq 1-A\sum {\rm eig}\, \partial v_i/\partial q_j$ (where A is a positive constant, see also \citealt{Kitaura2012}). The reversal after shell crossing must thus come from the 'collective' term. Care has to be taken when comparing these results to velocity fields which are sampled on a mesh and then differentiated on that mesh, since in that case the discontinuous velocity jumps across caustics completely dominate the differential fields so that one essentially only measures the motion of caustics. The exact derivatives that can be computed using the triangulated sheet do not suffer from these inherent problems.

\begin{figure}
\begin{center}
 \includegraphics[width=0.7\textwidth]{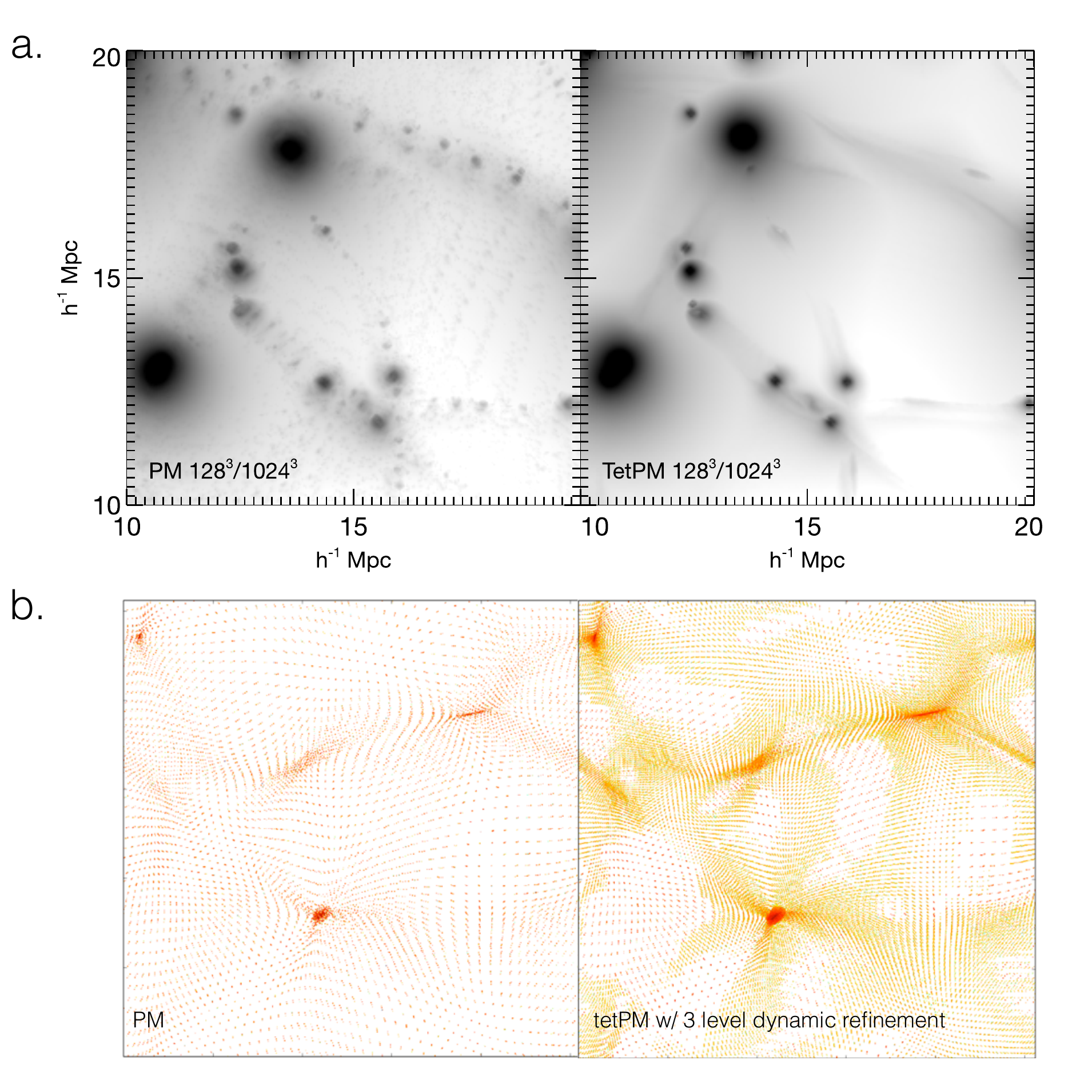} 
 \caption{{\bf Top:} Magnitude of the gravitational force in a simulation of structure formation in a warm dark matter cosmology. The N-body method (top left) shows artificial fragmentation of the filaments into clumps due to inherent discreteness effects. Spreading the mass between particles using tetrahedral elements strongly suppresses such discreteness effects. {\bf Bottom:} Adaptive refinement of the phase space sheet (right) will allow to track its rapid super-Lagrangian growth in the inner of halos.}
   \label{fig5}
\end{center}
\end{figure}

Another exciting possibility is to attempt to evolve the N-body system self-consistently using the tetrahedral approximation to the distribution function \citep{Hahn2013}. In such an approach, the vertices become tracers of the flow and the mass is, unlike in the N-body approach, not centered at the vertices, but is contained in the tetrahedral volume element. \cite{Hahn2013} have shown that this ``TetPM'' approach significantly reduces discreteness effects known for N-body simulations \citep[e.g.][]{Wang2007}. We show the result of such a simulation, where the mass distribution due to the tetrahedra is approximated by sampling the tetrahedra with mass-tracing pseudo-particles, in the top panels of Figure~\ref{fig5}. The top left panel shows the well known artificial fragmentation of filaments in warm dark matter simulations if the N-body method is used at a high force to low mass resolution ratio. The top right panel shows that the tetrahedral approach spreads the mass smoothly between the flow tracers leading to sharp well-defined filaments that show no sign of fragmentation. 

These fragments have made it inherently difficult if not impossible to measure the mass function of halos in warm dark matter (WDM) cosmologies robustly \citep[e.g.][]{Schneider2013} and their very presence might well invalidate many other predictions from such simulations. Using the TetPM method \cite{Angulo2013} were able to measure the WDM halo mass function in the absence of such fragments and found a distinctly different picture of large-scale structure and halo formation than in CDM: smooth pancakes permeated by filaments with very dense cores connecting the nodes of the cosmic web. They found that halos do not form below the truncation scale of the power spectrum, but come only into existence once the last axis collapses, quite in accord with the prediction of Zeldovich's equation~(\ref{eq:density}).

The rapid (phase and possibly chaotic) mixing of dark matter as it orbits in the potential wells of dark matter halos leads to a growth of the dark matter sheet that cannot be tracked by the Lagrangian motion of the tracer particles alone. The approximation that the tetrahedra describe the neighbourhood of a Lagrangian point $\mathbf{q}$ well breaks down as a consequence. The practical effect of this is that mass gets preferentially assigned to the center of halos making them more and more dense \citep{Hahn2013}. This limits the applicability of a tessellation approach based on a fixed number of Lagrangian tracers in the inner parts of dark matter halos where strong mixing occurs. This problem can however be resolved by inserting new vertices in the tessellation, i.e. by splitting the tetrahedra adaptively, in order to keep following the evolution of the dark matter sheet through all its complicated foldings at late times. The effect of such a refinement can be seen in the bottom panels of Figure~\ref{fig5}. In the left panel a standard N-body particle distribution is shown, while the right panel shows the distribution of flow tracers after tetrahedra were allowed to split into smaller units. The newly inserted vertices clearly trace out the regions of strongest deformation of the sheet. An article discussing the feasibility of such a adaptive dynamic refinement technique of the tessellation is currently in preparation (Hahn and Angulo 2015). Refinement will allow to track the full evolution of the fine-grained distribution function in the deeply non-linear regime.

\section{Summary}
\noindent In this article, we have attempted to present an overview of the connection between Zeldovich's groundbreaking work from 1970 describing the evolution of Lagrangian fluid elements in an expanding universe \citep{Zeldovich1970},  predicting a hierarchy of structures of pancakes, filaments and clusters, and modern work to dissect the cosmic web into distinct components. We have described how this work has inspired methods to classify the large-scale structure into dynamically distinct structure through the signature of eigenvalues of the tidal tensor or the velocity deformation tensor. Arguably the most interesting result of these attempts has been the discovery of (mass-dependent) alignments of the spin and shape of dark matter halos with the surrounding cosmic web. 

In the second part, we described how Zeldovich's description of the Lagrangian motion of fluid elements and their tidal deformation is equivalent to recent attempts to decompose the dark matter sheet (or Lagrangian manifold) into finite volume elements with the help of tessellations. The resulting tetrahedra, spanned by four tracer particles (which can be the N-body particles of a cosmological simulation) are finite difference approximations to Zeldovich's famous formula and describe the same dynamical behaviour allowing for shell-crossing along three distinct directions. We gave a brief overview of the published applications of these tessellation approaches to a finite volume Zeldovich approximation in terms of using them to provide new insights into the phase space structure of N-body simulations. Finally we discussed how they can be used to create alternative simulation methods for self-gravitating dark matter in the cold limit that evolve the tessellated phase space sheet self-consistently in phase-space. The rapid wrapping of the phase space sheet inside of halos where mixing occurs necessitates the adaptive refinement of the tessellation. Adaptively refined tesselations are a method which is currently in development and will provide new exciting insights into structure formation since the adaptive refinement allows to follow the full evolution of the fine-grained distribution function of dark matter as it is warped through gravitational collapse from pancakes to filaments and finally folded up in a halo. Information which is unaccessible for N-body methods and can only approximately be reconstructed without dynamic refinement.

\section*{Acknowledgments}
\noindent OH acknowledges support from the Swiss National Science Foundation through the Ambizione fellowship. The author thanks Tom Abel, Raul Angulo, Cristiano Porciani and Ralf Kaehler for many discussions surrounding the topics of this article, as well as Sergei Shandarin, Rien van de Weygaert, Enn Saar and Jaan Einasto for the invitation to speak at such a wonderful conference on the occasion of Zeldovich's 100th birthday.

\end{document}